\documentclass[aps,prl,twocolumn,superscriptaddress,amsmath]{revtex4-1}
\usepackage{graphicx}
\usepackage{natbib}
\usepackage[breaklinks]{hyperref}

\begin{document}
\title{Strong thermomechanical squeezing via weak measurement}

\author{A. Szorkovszky}
\author{G.A. Brawley}
\affiliation{Centre for Engineered Quantum Systems, University of Queensland, Australia}
\author{A.C. Doherty}
\affiliation{Centre for Engineered Quantum Systems, University of Sydney, Australia}
\author{W.P. Bowen}
\email[]{wbowen@physics.uq.edu.au}
\affiliation{Centre for Engineered Quantum Systems, University of Queensland, Australia}

\begin{abstract}
We experimentally surpass the 3dB limit to steady state parametric squeezing of a mechanical oscillator. The localization of a AFM cantilever, achieved by optimal estimation, is enhanced by up to 6.2 dB in one position quadrature when a detuned parametric drive is used. This squeezing is, in principle, limited only by the oscillator Q-factor. Used on low temperature, high frequency oscillators, this technique provides a pathway to achieve robust quantum squeezing below the zero-point motion. Broadly, our results demonstrate that control systems engineering can overcome well established limits in applications of nonlinear processes. Conversely, by localizing the mechanical position to better than the measurement precision of our apparatus, they demonstrate the usefulness of nonlinearities in control applications.
\end{abstract}

\maketitle

High-quality mechanical oscillators are widely used for weak force detection\cite{force1,force2}, nano-scale manipulation\cite{nems1,nems2} and quantum state engineering\cite{quantum1,quantum2}. Such applications often utilize optimal estimation to localize the oscillator, followed by feedback control to confine its position. In a classical context, this type of control is commonly used to linearize the response of sensors driven into their nonlinear regime, resulting in increased dynamic range and suppression of resonance frequency fluctuations\cite{hall}. Furthermore, spurred by the growing prospect of accessing new quantum physics\cite{blencowe}, similar techniques are now being applied to state-of-the art mechanical oscillators to cool them close to the quantum limit set by mechanical zero point motion\cite{feedback1,feedback2}, and ultimately surpass it via quantum control techniques such as back-action evasion\cite{clerk,vanner,hertzberg}. However, the level of achievable oscillator localization has always previously been limited to at best the measurement precision, presenting a significant barrier to applications in both quantum and classical regimes.

\begin{figure}[!t]
\centering
\includegraphics[width=8cm, bb=61 203 529 720]{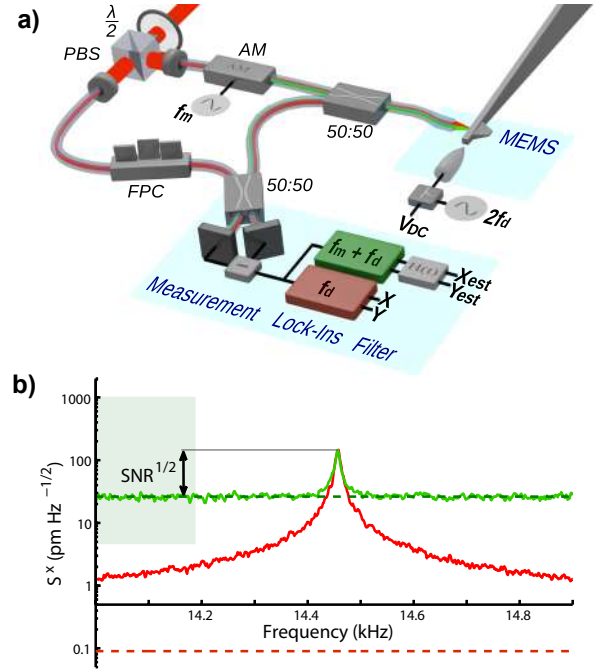}
\caption{\label{fig1}a) Schematic of the experimental setup. The red path in the fiber-based interferometer denotes the high-fidelity carrier signal while the green path denotes the low-fidelity signal created by amplitude modulation (AM). PBS: polarising beam-splitter; FPC: fiber polarisation controller. b) Displacement noise spectrum around the fundamental cantilever resonance measured by the laser carrier (red) and by the sideband created with a $1\mathrm{V}_\mathrm{pp}$ modulation (green). Dotted lines represent the respective shot-noise limited sensitivities of the measurements, while the green band corresponds to the range of sensitivity available from the utilised sideband intensities.}
\end{figure}

Applications of mechanical oscillators can also benefit from nonlinearities without requiring any measurement. An example of particular relevance to this article is mechanical parametric amplification, where direct modulation of the spring constant induces amplification of in-phase motion\cite{rugar}. This technique is often used in micro- and nano-electromechanical systems (MEMS/NEMS) to boost mechanical signals from in-phase forces above the measurement noise floor\cite{paramp1,paramp3}. Conversely, out-of-phase motion is deamplified and thereby more strongly confined, with its variance said to be "squeezed". In principle, squeezing below the zero-point motion variance $V_g$ is possible. Such ``quantum squeezing'' has applications in quantum metrology and tests of macro-scale entanglement and quantum gravity\cite{blencowe}. However, the emergence of mechanical instability limits the improvement in confinement to at most 50\%  (or -3dB) in the steady-state\cite{prl}. This 3dB limit impedes both classical and quantum applications of parametric de-amplification. For instance, since a mean thermal occupancy of just half a phonon increases the oscillator's motional variance to twice the zero-point motion variance, the 3dB limit imposes a strict pre-cooling requirement for quantum squeezing by this method. For typical micro- and nano-mechanical oscillators with resonance frequencies in the range of 1-100MHz, this temperature bound lies between $0.05$ and 5mK, outside the range of most conventional cryogenic setups.

Here, we combine control techniques and parametric modulation to both break the 3dB limit for the first time and achieve mechanical localization exceeding the measurement sensitivity of our apparatus by 6.2dB. The key concept, proposed recently in Ref.~\cite{prl}, is to induce correlations between the amplified and squeezed motional quadratures by detuning the parametric modulation. Information encoded on the amplified quadrature then allows the squeezed quadrature to be estimated with enhanced precision. Our experiments are performed with a conventional AFM cantilever at room temperature, and as such are far from the quantum regime. The enhanced localization possible through such ``thermomechanical squeezing''  can, however, be useful in force measurement; for instance, by increasing the dynamic range when signal distortion is introduced at large amplitudes\cite{hall,ion1}; by broadening the bandwidth in the squeezed quadrature\cite{mertz}; and by enhancing the sensitivity to pulsed forces with known timing\cite{vitali,weld}. Furthermore, since the technique demonstrated here applies equally to quantum zero-point noise, it provides a path towards precise quantum control and robust quantum squeezing of mechanical oscillators at attainable temperatures and in the absence of strong measurement\cite{prl}.

The experimental setup, shown in Fig.\ \ref{fig1}a, is based on a commonly used optical measurement of the mechanical element in a typical MEMS. The position of a gold-coated AFM cantilever is monitored using a Mach-Zender interferometer in a balanced homodyne configuration, with a fiber tip used to focus the optical field onto the cantilever. The sensitivity $S_x$ of the interferometer is 90$\mathrm{fm}/\sqrt{\mathrm{Hz}}$, as shown in Fig.\ \ref{fig1}b. This allows a high-fidelity measurement of the thermal noise in the fundamental mode of the cantilever, which is used to characterise its motion and the accuracy of our estimation procedure. The measurement noise for this signal $V_{\mathrm{meas}}=4\gamma S_x^2$, where $\gamma$ is the mode's decay rate, is 60dB below the thermally-induced variance $V_T$. A weak sideband is also created using an intensity modulation of the bright field, providing a low-fidelity measurement with independent shot-noise characteristics which was used to perform position estimation. The sensitivity of this low-fidelity measurement could be varied between 25 and 1000$\mathrm{pm}/\sqrt{\mathrm{Hz}}$ by adjusting the optical modulation depth, as illustrated in Fig.\ \ref{fig1}b. At room temperature, the thermal noise signal lies within this region, allowing the study of estimation techniques in the important regime where the signal level is comparable to the measurement noise floor, i.e.\ where the signal-to-noise ratio $\mathrm{SNR}=V_T/V_{meas}\approx 1$.

\begin{figure}[!t]
\centering
\includegraphics[width=8cm, bb=50 166 786 925]{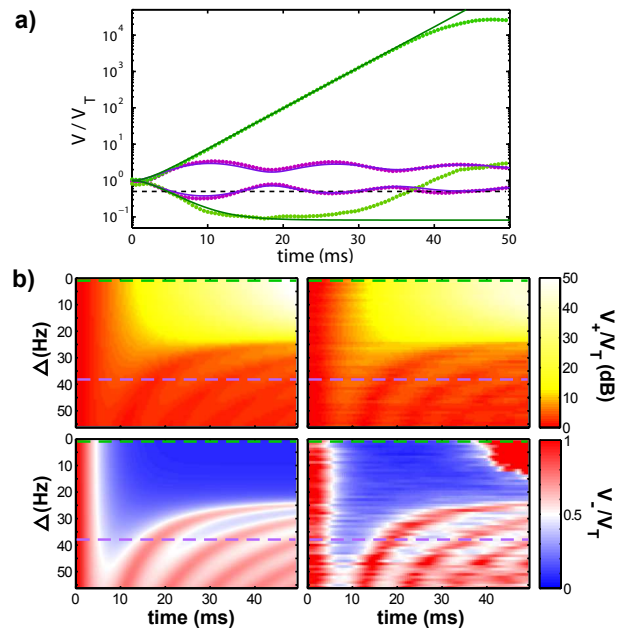}
\caption{\label{fig2}Evolution of the squeezed and antisqueezed quadratures with a continuous parametric drive of strength $\chi=22.5\mathrm{Hz}$ turned on at $\mathrm{t=0}$ and $f_0=12.5\mathrm{kHz}$. a) Normalised quadrature variances vs time for an on-resonance drive (green) and for a below-threshold detuned drive with $\Delta=38\mathrm{Hz}$ (violet). Solid lines are theoretical fits, while points show experimental statistics generated from 200 iterations of the drive turn-on. The dotted line represents the -3dB steady-state squeezing limit. At each point in time, the quadratures are rotated so that the covariance $\langle XY \rangle - \langle X\rangle \langle Y\rangle\approx 0$ over all iterations. A ring-up time of $2.5\mathrm{ms}$ is chosen for the parametric drive to minimise impulse forces on the cantilever. b) Theoretical (left) and experimental (right) variances as a function of detuning and time. Blue areas indicate squeezing below 3dB.}
\end{figure}

Parametric amplification requires the spring constant (and hence frequency) of the oscillator to be modulated near twice the resonance frequency $f_0$. For our cantilever, the spring constant was increased well above its intrinsic value $k_0\approx0.06\mathrm{N/m}$ by applying 450-650 volts between the cantilever and a nearby electrode; an effect due to the nonlinear position dependence of capacitive energy in this geometry\cite{rugar}. Since the frequency shift is proportional to the square of the voltage, this DC offset increases the peak-to-peak frequency modulation $\chi$ of the oscillator's fundamental mode from an additional alternating voltage\cite{paramp1}. The position measurements were fed into lock-in amplifiers with a bandwidth much wider than the mechanical decay rate $\gamma$, allowing the position dynamics around $f_0$ to be observed in a rotating frame at a nearby reference frequency $f_d = f_0 + \Delta$. The lock-in outputs $X$ and $Y$, describing orthogonal quadrature components of the position $x=X\cos(2\pi f_d t)+Y\sin(2\pi f_d t)$, allow easy visualization of the amplitude and phase of oscillation. The parametric effect from an alternating voltage at frequency $2f_d$ applied between the electrode and the cantilever leads to a preferred phase of mechanical oscillation and hence a squeezed thermal distribution in the X-Y plane.

\begin{figure*}[ht]
\centering
\includegraphics[width=14cm]{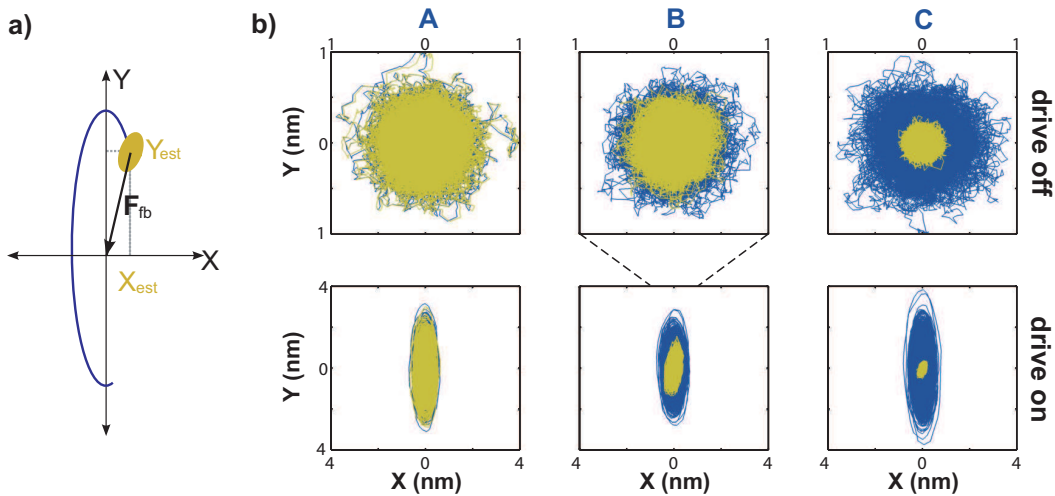}
\caption{\label{fig3}Reducing variance via estimation. a) Typical phase-space trajectory over a short time with a detuned parametric drive applied to the oscillator. The estimates $\{X_{est},Y_{est}\}$ at a given time are calculated by filtering the low-fidelity data, agreeing with a high-fidelity measurement (blue curve) to within an uncertainty given by the yellow ellipse and localizing the oscillator to this phase space region. A feedback force $F_{fb}$ confining the oscillator to near the origin can be modelled by subtracting the estimates from the high-fidelity data. b) Quadrature phase-space trajectories for 22.5 second samples obtained from high-fidelity measurement (blue) and the residual after subtracting the estimate (yellow). The upper panels show the random-walk pattern in the undriven case mixed down at the resonance frequency $f_0=14.5\mathrm{kHz}$ for weak (A), intermediate (B) and strong measurement (C). The lower panels show the elliptical trajectories and residual noise for a parametric drive strength of $\chi=57\mathrm{Hz}$ detuned close to threshold and using the same SNR as above.}
\end{figure*}

Initially, the high-fidelity measurement was used to analyse above and below-threshold position statistics under parametric amplification. Here, the DC voltage was set to 450V, shifting the fundamental mode frequency $f_0$ from 9.6kHz to 12.5kHz. When driven on resonance ($\Delta=0$), and with a strength above the instability threshold ($\chi>\gamma$), thermomechanical squeezing can surpass 3dB while the orthogonal quadrature is amplified indefinitely. Figure \ref{fig2}a shows the time-evolution of the maximally squeezed and antisqueezed quadrature variances measured in this regime, with $\chi=22.5\mathrm{Hz}$ and $\gamma=2\mathrm{Hz}$. Good agreement with theory is observed at short times ($<$20ms) with exponential growth of the amplified quadrature and thermomechanical squeezing approaching 11dB in the orthogonal quadrature. However, the amplified quadrature saturates after approximately 35ms, where the amplitude approaches the optical quarter-wavelength of 195nm and oscillations are no longer confined to the linear portion of the interference fringe. Crucially, a side-effect of this measurement nonlinearity is a severe degradation in observed squeezing well before saturation is apparent. Such limits to dynamic range therefore preclude the generation of all but transient squeezing above threshold. Nonetheless the strong squeezing observed reproduces the nonequilibrium squeezing observed in trapped ions\cite{ion2} for the first time in a micromechanical oscillator, albeit in the classical regime. Transient squeezing of this kind could be useful in applications where operation outside of equilibrium is acceptable, such as stroboscopic sensing\cite{vitali}.

By detuning the parametric drive off resonance, the oscillator phase undergoes a net rotation with respect to the amplification axis, increasing the instability threshold to $\chi_{th}=\sqrt{\Delta^2+\gamma^2}$\cite{njp}. Consequently for the drive strength used here ($\chi=\mathrm{22.5Hz}$), the oscillator is unstable for detunings below $\Delta=\mathrm{22.4Hz}$. When the detuning is increased further so that $\Delta>\chi$, the phase-space trajectories form stable elliptical orbits. The variances initially mirror this oscillatory behaviour before relaxing to steady-state values in the long time limit, in quantitative agreement with theoretical modelling\cite{njp}. The effect of increasing detuning on the transient statistics can be seen in Fig.\ \ref{fig2}b, where a dramatic change from monotonic behaviour to clear oscillations in the variance occurs at the threshold detuning. Notably, transient squeezing below 3dB is still possible below threshold, owing to the rapid drive turn-on. The final steady-state variances can be expressed with respect to the thermal variance $V_T$ as $V_\pm/V_T = (1\mp\chi/\chi_{th})^{-1}$\cite{njp}. The squeezing limit of $V_T/2$ is therefore --- while not a constraint outside of equilibrium --- a fundamental one when in the steady-state.

From the above observations, the benefit of a parametrically driven system in equilibrium would appear to be limited to enhanced readout in one quadrature and 3dB reduced variance in the other. However, it has been recently predicted that these phenomena can be combined to enhance localization using a weak measurement and optimal estimation\cite{prl}. For an oscillator detuned so that $\Delta>\chi$ and which has relaxed to the steady-state, the thermally excited oscillations will alternate between amplified and squeezed quadratures before decaying. Since the dynamics of the system are well known, a measurement of the amplified quadrature will provide some capacity to estimate the squeezed quadrature at a later time. The squeezed quadrature therefore obtains an effective sensitivity enhancement without amplifying its mechanical fluctuations. This is useful for localizing an oscillator where conditions such as cold environment, poor measurement sensitivity or high oscillator frequency limit the SNR.

Using control theory, the filter that extracts the best quadrature estimates from noisy measurements can be written in terms of parameters defining the oscillator's expected time-evolution along with the measurement sensitivity\cite{njp}. For an oscillator with no parametric drive, the optimal quadrature estimates $X_{est}$ and $Y_{est}$ that be obtained from the low-fidelity measurement records $\tilde X$ and $\tilde Y$ are of the form $X_{est}(t) = g_0\tilde X(t) * e^{-\Gamma_0 t}$ and $Y_{est}(t) = g_0\tilde Y(t) * e^{-\Gamma_0 t}$, where $*$ denotes a time convolution. With a detuned parametric drive turned on, the two quadratures become correlated and the estimates require a more complex convolution\cite{njp}, the implementation of which is described in the Supplementary Information. The best effective localization of the oscillator from these estimates can be quantified using conditional variances. For example, the conditional $X$ quadrature variance $V_X$ is the mean-square of the residual noise $X(t)-X_{est}(t)$. The conditional variance also defines the minimum effective temperature of a quadrature achievable from ideal feedback cooling, equivalent to applying phase-space displacements in the $X$ and $Y$ quadratures by $X_{est}$ and $Y_{est}$ respectively over time as illustrated in Fig.\ \ref{fig3}a.

Steady-state estimation was performed with the cantilever tuned to $14.5$kHz by a 650V bias and varying the sideband intensity to tune the SNR of the low-fidelity measurement. For each SNR, continuous low-fidelity measurements of the two quadratures of cantilever motion were recorded in both the undriven case and with an applied parametric drive of strength $\chi=57\mathrm{Hz}$ and detuning $\Delta=63\mathrm{Hz}$ to ensure the below-threshold condition. In both cases, optimal estimates $X_{est}$ and $Y_{est}$ were generated in post-processing by minimizing the respective conditional variances over the filter parameters.  Phase-space Brownian trajectories $\{X,Y\}$ determined from the high-fidelity measurement are plotted in Fig.\ \ref{fig3}b, along with corresponding residual noise $\{X\!-\!X_{est},Y\!-\!Y_{est}\}$ after applying the optimized filter to the low-fidelity measurements in the low, intermediate and high SNR regimes. As expected, when no parametric drive is applied, the quadratures of motion have equal uncertainty, determined by the optimal conditional variance i.e.\ $V_X=V_Y=V_0$.  As the thermal signal increases towards the noise floor (SNR approaches 1) the conditional variances drop sharply as expected. At maximum sideband intensity, the RMS uncertainty in both quadratures is reduced from the thermal value of 240pm to 60pm, corresponding to an effective temperature decrease from 300K to 19K. With the drive turned on, the high-fidelity measurement shows unconditional thermomechanical squeezing close to, but not surpassing, the 3dB limit. Critically, elliptical trajectories can be observed, establishing the correlations required for our estimation protocol between squeezed and antisqueezed quadratures. After subtracting the optimal estimate, the residual noise is maximally squeezed at an angle $\alpha$ that increases with SNR. The variance $V_\alpha$ of this quadrature decreases monotonically along with the antisqueezed variance as the measurement improves, with the residual noise becoming symmetric in the high SNR limit. 

\begin{figure}[!t]
\includegraphics[width=8.5cm, bb=18 0 870 362]{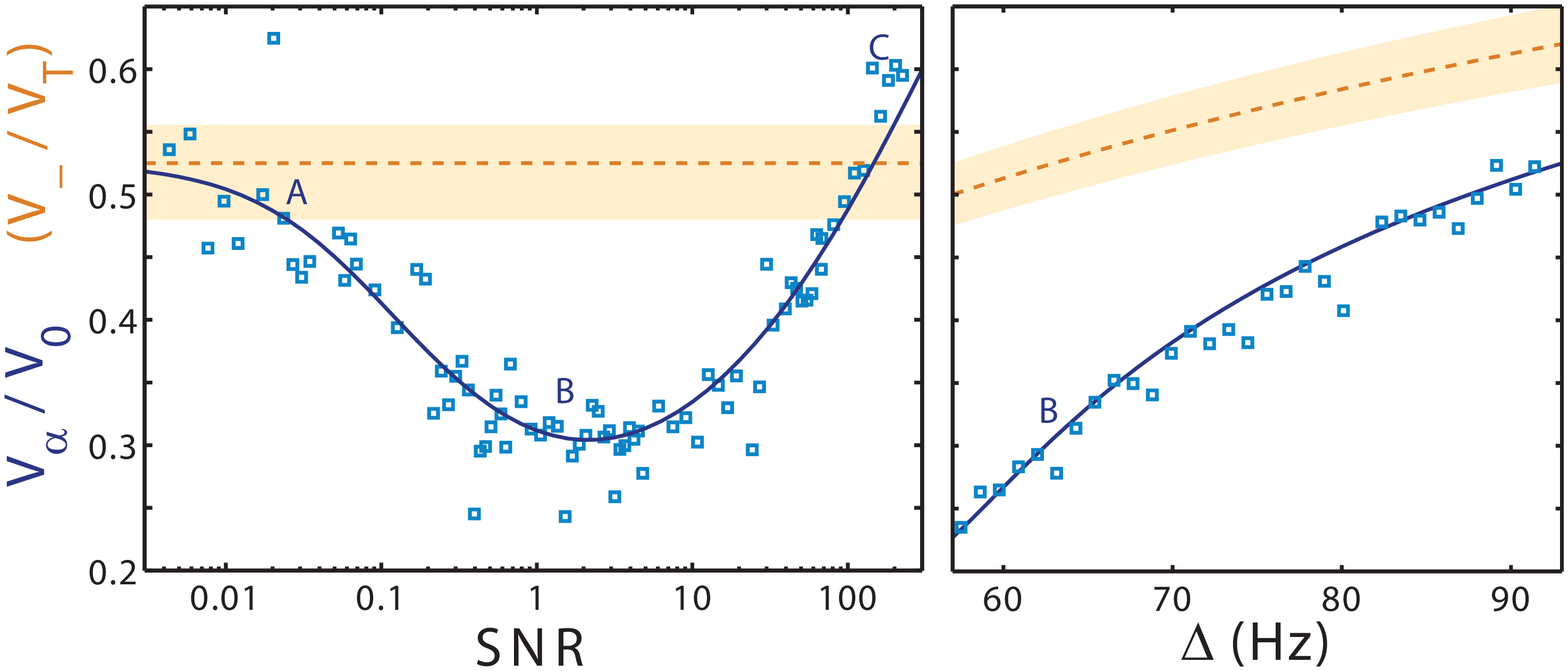}
\caption{\label{fig4}Steady-state squeezing using optimal estimation. Squeezing ratios are plotted against SNR for $\Delta=63\mathrm{Hz}$ (left) and against detuning for $\mathrm{SNR}\approx1$ (right). Blue points show the squeezing ratio, with theoretical fits shown as solid lines. Dotted red curves are fits to the squeezing without estimation, limited to 1/2, with shaded bands to represent the experimental error margin. Labels A-C indicate the datapoints used to generate the trajectories in Figure \ref{fig3}.}
\end{figure}

The squeezing ratio $V_{\alpha} / V_0$ determined from this analysis is shown in Fig.\ \ref{fig4} as a function of SNR, agreeing well with theory. As expected, the variances reproduce the unconditional squeezing in the weak measurement limit and the parametric drive has no effect in the strong measurement limit. However, in the intermediate regime where $\mathrm{SNR}\approx1$ there is a distinct minimum, allowing enhanced localization and breaking the 3dB limit by a significant factor. As can be seen in Fig.\ \ref{fig4} (right), the squeezing can be improved further by adjusting detuning closer to threshold, with a maximum thermomechanical squeezing of $6.2\mathrm{dB}$ achieved. These results can be understood by the fact that the effective increased sensitivity due to the parametric drive is of greatest benefit near the noise floor and with maximal amplification of the orthogonal quadrature. Since the maximum squeezing is proportional to $\sqrt{\chi/\gamma}$\cite{njp}, it can be enhanced by increasing the parametric drive strength, subject to the condition $\chi\ll f_0$. In principle, this allows arbitrary suppression of one quadrature of motion, exceeding the usual limit for control systems defined by the measurement precision. For applications requiring confinement in addition to localization, optimal estimates must be calculated in real-time in order to be fed back as a damping force. As shown here by optimisation, this can be achieved by using the well-defined filter parameters in Ref.\ \cite{njp}.

Although demonstrated with thermal fluctuations, our technique applies in the same manner to the zero-point motion of an oscillator, with the maximum reduction in conditional variance $V_{\alpha} / V_0$ independent of temperature\cite{njp}. The effect of the quantum modification at low temperatures---known as backaction noise---is instead to limit the initial conditional variance $V_0$ to be no lower than the ground state variance. Therefore our approach could enable strong quantum squeezing and ultra-precise quantum control\cite{wiseman}. Experiments with sensitivity near the standard quantum limit (where $\mathrm{SNR}\approx 1$ at zero temperature) have been recently performed with mechanical oscillators\cite{hertzberg,schliesser}. While purely measurement-based schemes exist to create mechanical squeezed states\cite{clerk}, significant squeezing requires high measurement strength and efficiency such that $\mathrm{SNR}\gg (2N_{th}+1)^2$. This regime is yet to be demonstrated in mechanical oscillators. Nanoelectromechanical systems are, however, commonly integrated with a parametric drive and can be pre-cooled to near the ground state\cite{schneider}. Such emerging systems are therefore good candidates for quantum squeezing below the zero-point motion using our technique, even without significant advances in transduction.

We have observed parametric thermomechanical squeezing of a micromechanical oscillator exceeding 3dB for the first time, both transient and in equilibrium, with the latter breaking a well-known limit for parametrically driven systems. This result demonstrates that the 3dB limit to steady-state parametric squeezing is not fundamental, and facilitates the wider use of thermomechanical squeezing in control and sensing applications.  The combination of parametric driving, measurement and estimation sheds light on the important interface between quantum measurement and control that is being approached most notably in opto- and electromechanical systems. The techniques introduced, if applied in conjunction with state-of-the-art readout techniques and high quality oscillators, also open the door for the engineering of nonclassical states of mesoscopic mechanical systems. More broadly, our results demonstrate that combining oscillator nonlinearity with control can both overcome fundamental limitations on parametric processes, as well as localize mechanical motion beyond constraints imposed by the measurement sensitivity.

\begin{acknowledgments}
This research was funded by the Australian Research Council Centre of Excellence CE110001013 and Discovery Project DP0987146.
\end{acknowledgments}

\clearpage

\appendix

\section{Supplementary Information: Filtering of time-series steady-state data}

\begin{figure}[!th]
\centering
\includegraphics[width=8.5cm]{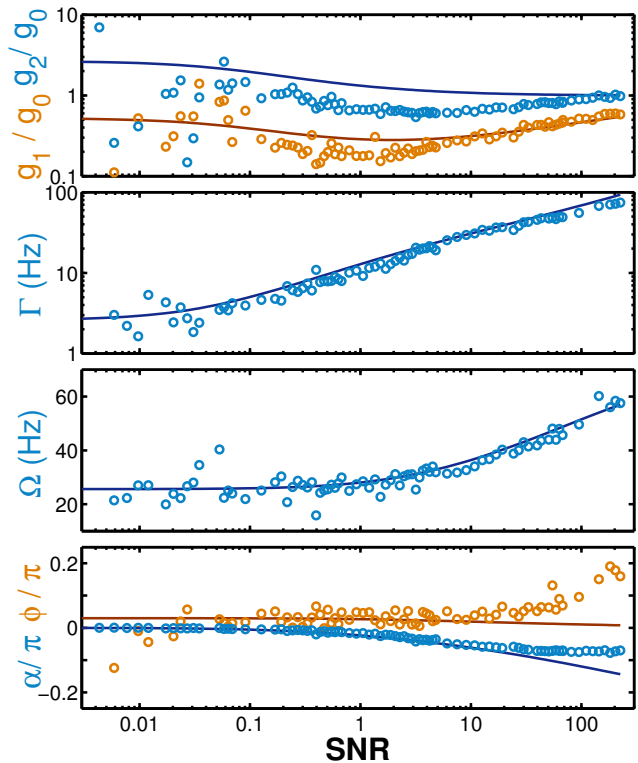}
\caption{\label{params}Optimal filter parameters at a constant near-threshold parametric drive. Solid lines indicate theoretical parameters for $\chi=57\mathrm{Hz}$,$\gamma=2.6\mathrm{Hz}$ and $\Delta=63\mathrm{Hz}$. Circles indicate parameters obtained from post-processing.}
\end{figure}

Optimal estimation was performed on quadrature measurements acquired while the cantilever was in thermal equilibrium. At each SNR of the low-fidelity measurement, simultaneous 45-second traces of the lock-in outputs $X,Y,\tilde X,\tilde Y$ were recorded using a data acquisition unit with no parametric drive applied to the cantilever, and with the fundamental mode frequency $f_0$ kept within 5Hz of the lock-in reference frequency. Data acquisition was then repeated using parametric drives of various detunings at the same SNR values.  The lock-in reference frequency --- now shifted by the parametric drive detuning $\Delta$ --- was kept phase-locked to the drive voltage. A lock-in time constant of $\tau_c=300\mu\mathrm{s}$ was used so that the output oscillations (limited in frequency to $\Delta$) were contained within the output bandwidth. 

In post-processing, the conditional variances $V_X$ and $V_Y$ were found by convolving the time series with a filter function (truncated to $22.5$ seconds) and calculating $\langle(X-X_{est})^2\rangle$ and $\langle(Y-Y_{est})^2\rangle$ over the second half of the data. The optimum conditional variance was then found by minimizing computationally over all filter parameters. The filter function, in both driven and undriven cases, has the general form

\begin{equation}\label{eq}
\left[\begin{matrix}
        X_{\alpha, est} \\
        Y_{\alpha, est}
       \end{matrix}\right]
 = \mathbf{H}(t)*\left[\begin{matrix}
        \tilde X_\alpha(t) \\
        \tilde Y_\alpha(t)
       \end{matrix}\right] \; ,
\end{equation}
and the optimal filter matrix $\mathbf{H}(t)$ takes the general form\cite{njp}
\begin{equation}\label{eq2}
\mathbf{H}(t) = \left[\begin{matrix}
        g_1\cos(\Omega t-\phi) & g_2\sin(\Omega t)\\
        g_3\sin(\Omega t) & g_4\cos(\Omega t+\phi)
       \end{matrix}\right]e^{-\Gamma t} \; .
\end{equation}
where $g_n$, $\Omega$, and $\phi$ are positive real numbers. This filter function, in the parametrically driven case, assumes that $\chi<\Delta$ and that the quadratures are rotated by an angle $\alpha$ such that $X_\alpha-\tilde X_\alpha$ is maximally squeezed.

In the undriven case, the filter is simplified by the restrictions $g_n=g_0$ (for all $n$), and $\alpha=\phi=0$. Here, the rotation frequency $\Omega$ is kept to account for drifts in $f_0$. The remaining parameters $\Gamma$ and $g_0$ are functions of SNR.

In the parametrically driven case, the measurement quadratures were initially rotated to be aligned with the unconditional squeezing. The additional rotation angle $\alpha$ for which maximum conditional squeezing occurs was included as an optimisation parameter. This angle, as well as the filter parameters, are functions of SNR, decay rate $\gamma$, drive strength $\chi$ and detuning $\Delta$. Initial estimates of the filter parameters were calculated based on these known quantities according to theory derived in Ref.\ \cite{njp}. 

Fig.\ \ref{params} compares the initial estimates of filter parameters to the converged optimal values used to obtain the results in Fig.\ \ref{fig4} (left). The optimised filter agrees reasonably well with theoretical estimates, although is more likely to converge to consistent values in the higher-fidelity regime and in regions with more distinct minima. This demonstrates that the optimal filter parameters described in Ref.\ \cite{njp} agree quantitatively with experiment.

\end{document}